\documentstyle[11pt,epsfig]{article}
\begin{document}
\thispagestyle{empty}
\setcounter{page}0

\begin{flushright}
CERN-PH-TH/2004-113
\end{flushright}

~\vfill
\begin{center}
{\Large\bf On \boldmath $V_{ud}$ determination from neutron decay}

\vfill

{\large M. V. Chizhov}
 \vspace{1cm}

{\em
Theory Division, Department of Physics, CERN,\\
CH-1211 Geneva 23, Switzerland\\
and\\
Centre of Space Research and Technologies, Faculty of Physics,\\
University of Sofia, 1164 Sofia, Bulgaria
}

\end{center}  \vfill

\begin{abstract}

The recent results of the PIBETA experiment strongly suggest the
presence of a non-($V\!-\!A$) anomalous interaction in the
radiative pion decay. The same interaction should inevitably
contribute to the neutron decay and in particular it should affect
the $V_{ud}$ determination. This paper is dedicated to the
prediction of the eventual discrepancy in $V_{ud}$ extracted from
the super-allowed $0^+\to 0^+$ Fermi transitions and from the
polarized neutron decay.

\end{abstract}

\vfill

\newpage

\section{Introduction}
The standard model (SM) includes three generations of quark
doublets. However, our world consists of only $u$ and $d$ quarks,
the lightest flavours of the first generation. This occurs through
quark-generation mixings in charge-changing weak decays. The quark
mixings cannot be predicted within the SM and they are matter of
experimental investigations and theoretical speculations.

If there exist only three quark generations, then the transition
probability, for example, of $up$-quark $u$ to all $down$-quarks
$d$, $s$ and $b$ should be equal to one
\begin{equation}
\vert V_{ud}\vert^2+\vert V_{us}\vert^2+\vert V_{ub}\vert^2=1.
\label{unitarity}
\end{equation}
This unitarity condition can be tested experimentally.

Already in the 1990's~\cite{90} it was noted that there is a small
deviation from unitarity. Therefore, this issue is attracting
great interest and lots of experimental efforts were made to
understand this problem. At present experimental accuracy of
$V_{ud}$ and $V_{us}$ determination, the third matrix element
$V_{ub}$ can be safely neglected in (\ref{unitarity}), thanks to
its small value, and the problem is reduced to a verification of a
simple trigonometrical identity
\begin{equation}
\cos^2\vartheta_C+\sin^2\vartheta_C=1
\label{Cabibbo}
\end{equation}
for the Cabibbo angle $\vartheta_C$. Therefore, an independent and
self-consistent determination of the first two matrix elements is
of  great importance.

At present, after a first indication from the E865
Collaboration~\cite{E865} that the value of $V_{us}$ could be
higher than the PDG value 0.2196(23)~\cite{PDG}, other
experiments~\cite{KLOE,KTEV} also confirm this result. Reanalysis
of the hyperon beta decay~\cite{hyperon} also leads to a somewhat
higher value of $V_{us}\simeq \sin\vartheta_C=0.2250(27)$, which
is in better agreement with the unitarity condition
(\ref{Cabibbo}). This value surprisingly coincides with the value
0.2238(30)~\cite{ratio} determined from the ratio of experimental
kaon and pion decay widths $\Gamma(K\to\mu\nu)/$
$\Gamma(\pi\to\mu\nu)$~\cite{PDG} using the lattice calculations
of the pseudoscalar decay constant ratio
$f_K/f_\pi$~\cite{lattice} and assuming unitarity. All these facts
indicate a higher value of $V_{us}$  and hence the unitarity
problem does not exist.

Meanwhile, different experiments have been dedicated to
independent determinations of the first matrix element $V_{ud}$.
The most precise result, $V_{ud}\simeq\cos\vartheta_C=0.9740(5)$,
comes from a series of experiments on super-allowed $0^+\to 0^+$
Fermi transitions~\cite{Vud}. Recently, $V_{ud}=0.9713(13)$ has
been derived, on a comparable precision level, from the polarized
neutron decay~\cite{PERKEO}. A compatible but less precise result
$V_{ud}=0.9728(30)$~\cite{PIBETA} has been achieved by measuring
the rare pion beta decay.

Although the unitarity problem is certainly solved by now, another
problem is probably emerging in connection with a very precise
determination of the first matrix element from the super-allowed
$0^+\to 0^+$ Fermi transitions and from the polarized neutron
decay. The present measurements indicate a 2$\sigma$ difference
for the extracted $V_{ud}$. The situation can be clarified in the
future with new experiments for measurements of the angular
correlation coefficient $a$ and the asymmetry parameter $A$ in the
neutron beta decay at the sub-10$^{-3}$ level~\cite{CKMw}.

This paper is dedicated to the prediction of an eventual
discrepancy in $V_{ud}$ extracted from the super-allowed $0^+\to
0^+$ Fermi transitions and from the polarized neutron decay. The
key difference between these two methods is related to the fact
that polarization phenomena are very sensitive to chiral
structures other than $V-A$. Therefore, a possible new tensor
interaction, which explains the anomaly in the radiative pion
decay~\cite{discovery}, can be responsible for this discrepancy.

\vspace{0.5cm}

\section{\boldmath $V_{ud}$ determination from neutron decay}

A determination of the strength of the $u\leftrightarrow d$ quark
transition with respect to the pure leptonic
\mbox{$e\leftrightarrow\nu_e$} transition can be made using
absolute measurements, {\it e.g.}\ partial widths or (and)
lifetime. Since the neutron has only one decay mode, $n\to
p\,e\,\bar{\nu}\,(\gamma)$, precise measurements of its mean
lifetime are absolutely necessary. Nevertheless, the knowledge of
only this parameter cannot allow us to determine $V_{ud}$, because
in the SM the neutron lifetime $\tau_n$,
\begin{equation}
\tau_n^{-1}\propto|V_{ud}|^2~G^2_F \left(1+3|\lambda|^2\right),
\label{lifetime}
\end{equation}
depends also on the phenomenological parameter $\lambda$, the
ratio of the axial coupling constant to the vector coupling
constant.\footnote{For the sake of simplicity we neglect, in the
following, all effects of the radiative corrections and additional
terms of recoil order, {\it e.g.}\ weak magnetism. However, such
terms should be taken into account in precision measurements of
neutron decay~\cite{precision}.}

Fortunately, in the neutron decay several observables are
accessible to experiments, which also depend on the same parameter
$\lambda$. For example, the decay probability for a polarized
neutron can be written~\cite{Jackson} as
\begin{equation}
\frac{{\rm d}\Gamma}{{\rm d}E_e{\rm d}\Omega_e{\rm d}\Omega_{\bar{\nu}}}
\propto
1+a~\frac{\mbox{\boldmath $p$}_e\cdot\mbox{\boldmath $p$}_{\bar{\nu}}}
{E_e E_{\bar{\nu}}}+
\langle\mbox{\boldmath $\sigma$}_n\rangle\cdot
\left[A~\frac{\mbox{\boldmath $p$}_e}{E_e}
+B~\frac{\mbox{\boldmath $p$}_{\bar{\nu}}}{E_{\bar{\nu}}}
+D~\frac{\mbox{\boldmath $p$}_e\times\mbox{\boldmath $p$}_{\bar{\nu}}}
{E_e E_{\bar{\nu}}}\right],
\label{aABD}
\end{equation}
where $\langle\mbox{\boldmath $\sigma$}_n\rangle$ is the neutron polarization.
The correlation coefficients $a$, $A$, $B$ and $D$ in the SM are
given by the relations:
\begin{equation}
a=\frac{1-|\lambda|^2}{1+3|\lambda|^2},\hspace{0.5cm}
A=-2\frac{|\lambda|^2+{\rm Re}(\lambda)}{1+3|\lambda|^2},\hspace{0.5cm}
B=2\frac{|\lambda|^2-{\rm Re}(\lambda)}{1+3|\lambda|^2},\hspace{0.5cm}
D=2\frac{{\rm Im}(\lambda)}{1+3|\lambda|^2}.
\end{equation}
A non-zero value of the triple correlation coefficient $D$ would
indicate $T$-violation; however, in the SM its value is predicted
to be vanishingly small. Present experiments confirm this
statement at the level of about 0.1\%. Therefore, in the following
we will consider the parameter $\lambda$ to be a real constant.

Taking into account the current world average value for
$\lambda=-1.2695(29)$~\cite{PDG} we can estimate the sensitivity
of the correlation coefficients to this parameter
\begin{eqnarray}
\frac{\delta a}{a}&=&\frac{8\lambda^2}{(\lambda^2-1)(1+3\lambda^2)}
\frac{\delta\lambda}{\lambda}\simeq 3.6~\frac{\delta\lambda}{\lambda},
\label{a}\\
\frac{\delta A}{A}&=
&\frac{(1-\lambda)(1+3\lambda)}{(1+\lambda)(1+3\lambda^2)}
\frac{\delta\lambda}{\lambda}~\simeq ~4.1~\frac{\delta\lambda}{\lambda},
\label{A}\\
\frac{\delta B}{B}&=
&\frac{(1+\lambda)(1-3\lambda)}{(1-\lambda)(1+3\lambda^2)}
\frac{\delta\lambda}{\lambda}\simeq -0.1~\frac{\delta\lambda}{\lambda}.
\label{B}
\end{eqnarray}
The parameters most sensitive to $\lambda$ appear to be the
angular correlation coefficient $a$ and the asymmetry parameter
$A$. Therefore, in the following we will concentrate only on these
correlation coefficients.

The recent results of the PIBETA experiment on radiative pion
decay~\cite{RPD} strongly suggest the presence of non-($V-A$)
anomalous interaction~\cite{discovery}
\begin{equation}
{\cal L}_T =  -f_T\frac{G_F V_{ud}}{\sqrt{2}}~
\bar{u}\sigma_{\lambda\alpha}d~
\frac{4q_\alpha q_\beta}{q^2}~
\bar{e}\sigma_{\lambda\beta}(1-\gamma^5)\nu_e +{\rm h.c.},
\label{new}
\end{equation}
with dimensionless coupling constant $f_T\simeq 0.013$, where
$q_\alpha$ is the momentum transfer between quarks and leptons. It
is obvious that the same interaction will affect all observables
of the neutron decay: the lifetime (\ref{lifetime}) and the decay
distribution~(\ref{aABD}).

In order to apply this interaction to the neutron decay we estimate
the matrix element
\begin{equation}
\langle p|\bar{u}\,\sigma_{\alpha\beta}d|n\rangle=
g_T~\bar{p}\,\sigma_{\alpha\beta}n
\end{equation}
using the technique of ref.~\cite{Poblaguev} with $g_T\simeq
4.7(m_u+m_d)/(m_n+m_p)= 0.029(8)$ for the current quark masses
$m_u+m_d = 11.5\pm 3.4$ MeV at 1 GeV~\cite{PDG}. This leads to a
new matrix element in the neutron decay
\begin{equation}
{\cal M}_T=-F_T\frac{G_F V_{ud}}{\sqrt{2}}~
\bar{p}\,\sigma_{\lambda\alpha}n~ \frac{q_\alpha q_\beta}{q^2}~
\bar{e}\sigma_{\lambda\beta}(1-\gamma^5)\nu_e,
\label{Mt}
\end{equation}
with an effective coupling constant $F_T=4f_T g_T = 1.5(4)\times
10^{-3}$, which is an order of the recoil effects
$(m_n-m_p)/m_n\equiv\Delta/m_n\approx 1.4\times 10^{-3}$.

It is worth while to note that the matrix element (\ref{Mt})
consists of two different terms with opposite nucleon chiralities
\begin{eqnarray}
{\cal M}_T=\hspace{-0.3cm}&-&\hspace{-0.3cm}F_T\frac{G_F
V_{ud}}{4\sqrt{2}}~ \bar{p}_R\,\sigma_{\lambda\beta}n_L~
\bar{e}\sigma_{\lambda\beta}(1-\gamma^5)\nu_e
\label{MtL}\\
&-&\hspace{-0.3cm}F_T\frac{G_F V_{ud}}{\sqrt{2}}~
\bar{p}_L\,\sigma_{\lambda\alpha}n_R~ \frac{q_\alpha
q_\beta}{q^2}~ \bar{e}\sigma_{\lambda\beta}(1-\gamma^5)\nu_e.
\label{MtR}
\end{eqnarray}
The first term (\ref{MtL}) is the usual local tensor matrix
element, which has been used in the literature for testing a
possible effect of new interactions. The second term (\ref{MtR})
is a new non-local tensor matrix element originated from the
non-local quark interaction (\ref{new}), which has been
constructed so as to avoid the constraints from the ordinary pion
decay~\cite{MPL}.

\vspace{1cm}

\section{Unpolarized neutron decay}

Taking into account the new matrix element (\ref{Mt}), the
differential energy distribution of an unpolarized neutron decay
at rest reads
\begin{eqnarray}
\hspace{-0.4cm}\frac{{\rm d}^2\Gamma_0}{{\rm d}E_e{\rm d}q^2}
\hspace{-0.2cm}&\propto&\hspace{-0.2cm}
2E_e\left(E_{m}-E_e\right)\left(1+\lambda^2\right)
-\frac{q^2-m^2_e}{2}\left(1-\lambda^2\right)
\nonumber\\
&+&\hspace{-0.2cm}2 F_T \left[2E_e\left(E_{m}-E_e\right)\frac{m_e}{E_e}
-\frac{q^2-m^2_e}{q^2}m_e E_{m}\right]\lambda
\nonumber\\
&+&\hspace{-0.2cm}F_T^2\left\{2E_e\left(E_{m}-E_e\right)-\frac{q^2+m^2_e}{4}
-\left[E_{m}\left(E_{m}-2E_e\right)-\frac{m_e^2}{2}\right]\frac{m_e^2}{q^2}
-E^2_{m}\frac{m_e^4}{q^4}\right\}\!,
\label{G0}
\end{eqnarray}
where $E_m=(m^2_n-m^2_p+m^2_e)/(2m_n)\approx \Delta$ is the
maximum electron energy, and the squared momentum transfer
$q^2=\Delta^2-2m_nT$ is connected to the proton kinetic energy~$T$
or to the electron--antineutrino angular correlation $q^2=m^2_e+2(
E_e E_{\bar{\nu}}-\mbox{\boldmath $p$}_e\cdot\mbox{\boldmath
$p$}_{\bar{\nu}})$.

The first line in the r.h.s. of eq.\ (\ref{G0}) is the SM
contribution in the limit of the non-relativistic nucleons. The
second line presents the contribution from an interference between
the SM matrix element and the new tensor matrix element
(\ref{Mt}). As expected this contribution is proportional to
$\lambda$, because the tensor interaction interferes only with the
Gamow--Teller amplitude. It consists of two terms: the well-known
Fierz interference term and the new contribution from the tensor
interaction with opposite nucleon chiralities (\ref{MtR}). The
third line stems from the square of the new matrix element
(\ref{Mt}) and can be safely neglected thanks to the second-order
contribution of the small parameter $F_T$.

On the one hand angular correlation measurements have the
advantage that it is not necessary to deal with a polarized
neutron. On the other hand, since the antineutrino is not
registered, only indirect methods have been employed through a
detection of low-energy recoil protons. The present accuracy of
$a$ does not exceed 5\%, which corresponds to a worse $\lambda$
uncertainty $\Delta\lambda\simeq 0.0139$ than that extracted from
the asymmetry parameter $A$~\cite{PDG}.

Meanwhile, new experiments have been proposed
\cite{Baessler,Yerozolimsky} with an improved accuracy of
$a$-measurements, using different experimental methods. The aim of
the collaboration~\cite{Baessler} is to use the neutron decay
spectrometer $a$SPECT to improve the precision of $a$ by more than
an order of magnitude, relying on the method based on measurements
of the proton kinetic energy spectrum. However, even at this
precision, it will be impossible to detect the contribution of the
new matrix element (\ref{Mt}).

It is interesting to note that an integration over the electron
energy spectrum leads to a zero result for the  interference term
in the second line in eq.\ (\ref{G0}), because the two different
contributions cancel each other. Therefore, the new tensor
interaction (\ref{Mt}) does not distort the recoil proton spectrum
and does not contribute to the neutron lifetime. The results of
the experiment should correspond to the SM predictions.

A different situation may occur in the case of direct measurements
of $a$ by recording the spectrum of electrons emitted into a given
range of angles referred to the proton momentum
\cite{Yerozolimsky}. However, the expected 1\% accuracy in the
value of  $a$ will probably be insufficient to detect the effect
of the new tensor interaction (\ref{Mt}) due to small coupling
constant $F_T$.

\vspace{1cm}

\section{Decay of the polarized neutron}

The decay of the polarized neutron allows us to extract $\lambda$
only from the well-measured electron spectrum
\begin{equation}
\frac{{\rm d}\Gamma}{{\rm d}E_e}= \frac{{\rm d}\Gamma_0}{{\rm
d}E_e}+ \langle\mbox{\boldmath $\sigma$}_n\rangle\cdot
\frac{\mbox{\boldmath $p$}_e}{E_e} ~\frac{{\rm d}\Gamma_A}{{\rm
d}E_e}, \label{GE}
\end{equation}
where
\begin{eqnarray}
\frac{{\rm d}\Gamma_0}{{\rm d}E_e}&\propto&
E_e\left(E_m-E_e\right)\left(1+3\lambda^2\right)
\nonumber\\
&+&2m_e F_T\left[E_m-2E_e
+E_m\frac{m^2_e}{2b}\ln\left|\frac{a+b}{a-b}\right|\right]\lambda
\label{GE0}
\end{eqnarray}
can be obtained from eq.\ (\ref{G0}), neglecting the last term and
integrating over $q^2$. Here the functions $a=2E_e(E_m-E_e)+m^2_e$
and $b=2(E_m-E_e)|\mbox{\boldmath $p$}_e|$ are related to the
maximum $q^2_{\rm max}=a+b$ and the minimum $q^2_{\rm min}=a-b$ of
the squared momentum transfer.

The anisotropic distribution in the presence of the new tensor
interaction (\ref{Mt}) can be easily calculated:

\begin{eqnarray}
\frac{{\rm d}\Gamma_A}{{\rm
d}E_e}\propto\hspace{-0.2cm}&-&\hspace{-0.2cm}
2E_e\left(E_m-E_e\right)\left(\lambda^2+\lambda\right)
\nonumber\\
&-&\hspace{-0.2cm}\frac{m_e F_T E_e}{E_e^2-m_e^2}\left\{
E_e\left(E_m-E_e\right)+\frac{m^2_e}{2}-
\left[E_m\left(E_m-E_e\right)+\frac{m^2_e}{4}\right]
\frac{m^2_e}{b}\ln\left|\frac{a+b}{a-b}\right| \right.
\nonumber\\
&&\hspace{1.1cm}-\left.\left[E_e\left(E_m+E_e\right)-\frac{m^2_e}{2}
-\left(E^2_m-\frac{m^2_e}{4}\right)\frac{m^2_e}{b}\ln\left|\frac{a+b}{a-b}\right|
\right]\lambda\right\}, \label{GA}
\end{eqnarray}
where only the SM contribution and the leading-order term in
$F_T$ are shown as well. This new term stems only from the
interference between the SM matrix element and the tensor
interaction for the transition of the right-handed neutron to the
left-handed proton (\ref{MtR}). It is absent in the usual case of
the local tensor matrix element for the transition of the
left-handed neutron to the right-handed proton (\ref{MtL}).

The new contribution in eq.\ (\ref{GA}) is negative over the whole
electron spectrum and it leads effectively to a larger absolute
value of the asymmetry parameter $A$ than in the SM for the same
parameter $\lambda$. Therefore, to extract the right value of
$\lambda$, the experimental asymmetry should be fitted according
to eqs.\ (\ref{GE}), (\ref{GE0}), (\ref{GA}), taking into account
the contribution of the new tensor interaction.

To estimate the effect of this interaction we plot in fig.~1 the
ratio of the asymmetry parameter $A={\rm d}\Gamma_A/{\rm
d}\Gamma_0$ to its value $A_0$ at $F_T=0$ for the region of the
electron spectrum fitted by the PERKEO II
Collaboration~\cite{PERKEO}. It shows in average 0.7\% systematic
contribution from the tensor interaction, which is the same as the
accuracy of the experiment. Therefore, the real value of the
measured parameter $\lambda=1.2720(19)$ can be obtained from the
experimental value $\lambda^{\rm exp}=1.2739(19)$ by shifting it
down with the value of the experimental accuracy,
$\Delta\lambda=0.0019$.

\begin{figure}[th]
\mbox{\hspace{2cm}}\epsfig{file=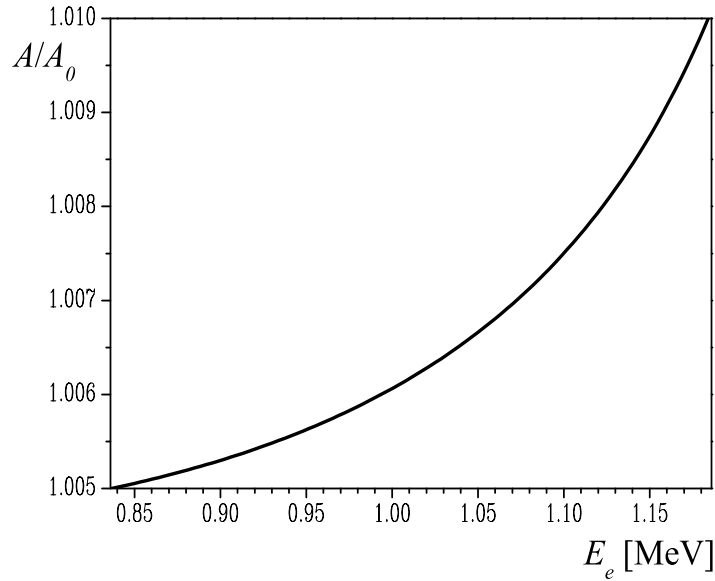,height=9cm,width=10.5cm}
\vspace{-0.5cm} \caption{The ratio of the asymmetry parameter
$A={\rm d}\Gamma_A/{\rm d}\Gamma_0$ to its value $A_0$ at $F_T=0$
for the region of the electron spectrum fitted by the PERKEO II
Collaboration.}
\end{figure}

Taking into account all effects from the radiative corrections and
the phase-space factor, and also the fact that the new tensor
interaction does not affect, in the leading order, the neutron
lifetime, eq.\ (\ref{lifetime}) can be rewritten
as~\cite{Marciano}
\begin{equation}
|V_{ud}|^2=\frac{4908\pm 4 s}{\tau_n\left(1+3\lambda^2\right)}.
\end{equation}
Using $\tau_n=885.7(8)s$ and $\lambda=1.2720(19)$, it leads to the
corrected value $V_{ud}=0.9729(13)$ extracted from the polarized
neutron decay, which is in a better agreement with
$V_{ud}=0.9740(5)$ extracted from super-allowed beta decays.

\section{Conclusions}

Recently strong evidence for a deviation from the SM has been
obtained by the PIBETA Collaboration~\cite{RPD}. Namely, the SM
fails to describe the energy distribution and the branching ratio
of the radiative decays of positive pions at rest in the
high-$E_\gamma$/low-$E_e$ kinematic region of the Dalitz plot. The
previous experiment, performed by the ISTRA
Collaboration~\cite{ISTRA}, testing the radiative decays of
negative pions in flight in a wide kinematic region, had announced
the same effect, although statistically less significant. The
present PIBETA result~\cite{RPD} indicates a deficit of the
branching ratio of the radiative pion decay in the specified
kinematic region at the $8\sigma$ level with respect to the SM
prediction, while in the other kinematic regions both the
branching ratios and the energy distributions are compatible with
the \mbox{$V\!-\!A$} interaction.

The anomaly observed by the ISTRA Collaboration has been explained
in the framework of an extended theory of the electroweak
interactions, with a new type of fundamental particles -- chiral
spin-1 bosons~\cite{MPL} described by antisymmetric second-rank
tensor fields. An exchange of these particles leads effectively to
the phenomenological tensor interaction (\ref{new}). The same
interaction should contribute to the neutron decay as well.
However, it has been shown here that with the present experimental
accuracy we cannot conclude definitely about their presence in the
neutron decay. The only hint of their manifestation is the partial
explanation of the about 2$\sigma$ discrepancy in $V_{ud}$
extracted from the polarized neutron decay and from the
super-allowed beta decays. Probably, new experiments under
construction, {\it e.g.\ } the UNCA and the abBA, using new
Spallation Neutron Source facilities and aiming at high-precision
measurements of the neutron decay parameters, could definitely
confirm or reject the predicted distortions of the spectrum due to
the new tensor interaction.

Based on the universality of couplings of the new chiral bosons to
leptons and quarks, the extended electroweak model~\cite{MPL}
predicts an admixture of analogous tensor interactions in pure
lepton processes~\cite{muon}. However, at present it is not easy
to detect them in the ordinary muon decay~\cite{TWIST} as well.
Meanwhile, the new tensor interaction (\ref{new}), with the same
effective coupling constant, allows us to explain~\cite{tau} the
painful widely commented discrepancy in the two-pion spectral
functions extracted from the $e^+e^-$ annihilation and from the
$\tau$-decay. It is hoped that future experiments will clarify
this situation better.

\section*{Acknowledgements}
The author acknowledges the warm hospitality of the Theory
Division at CERN, where this work has been fulfilled.

\pagebreak[4]

\end{document}